\documentclass{article}
\newcommand{\mi}{microcalcification }

\begin{document}
\title{
A MULTI-SCALE APPROACH TO THE COMPUTER-AIDED DETECTION OF MICROCALCIFICATION CLUSTERS IN DIGITAL MAMMOGRAMS
}
\author{
P. Delogu, M.E. Fantacci, A. Retico, A. Stefanini and A. Tata \\
{\em Dipartimento di Fisica dell'Universit\`a e INFN di Pisa,} \\
{\em Largo Pontecorvo 3, 56127 Pisa, Italy}\\
}
\date{}
\maketitle
\baselineskip=11.6pt
\begin{abstract}
A computer-aided detection (CADe) system for the identification of microcalcification clusters in digital mammograms has been developed. It is
mainly based on the application of wavelet transforms for image filtering and
neural networks for both the feature extraction and the classification procedures.
 This CADe system is easily adaptable to different databases. 
We report and compare the FROC
curves obtained on the private database we used for developing the
CADe system and on the publicly available MIAS database. The results achieved on the two databases show the same trend, thus
demonstrating the good generalization capability of the system.
\end{abstract}
\baselineskip=14pt
\section{Introduction}
Microcalcifications appear as small bright circular
or slightly elongated spots   embedded
in the complex normal breast tissue imaged in a mammogram.
Especially when they are grouped in clusters, microcalcifications
can be an important early indication of breast cancer. 
Computer-aided detection (CADe) systems can improve the radiologists' accuracy in the interpretation of mammograms by alerting them to suspicious areas of the image containing possibly pathological signs.

The main problem one has to deal with, in developing a CADe system for mammography,
is the strong dependence of the method, of the parameters and of the
performances of the system on the dataset used in the set-up and testing procedures.
The approach we adopted for our CADe system is mainly based on the exploitation of the properties
of the wavelet analysis and the artificial neural networks.
The use of wavelets in the pre-processing step, together
with the implementation of an automatic neural-based procedure for the feature
extraction, allows for a plan generalization of the analysis scheme 
 to databases characterized by different acquisition and storing parameters.
\section{CADe scheme}
The CADe  scheme can be summarized in the following main
steps:
\begin{itemize}
\item INPUT: digitized mammogram;
\item Pre-processing of the mammogram:
 identification of the breast skin line and  
segmentation of the breast region with respect to the background; 
 application of a wavelet-based filter in order to enhance 
the \mi signal; 
\item Feature extraction:
decomposition of the breast region  in several $N$$\times$$N$ 
pixel-wide 
sub-images to be processed each at a time;
automatic extraction of the features from each  sub-image; 
\item Classification: 
clustering of the processed sub-images  into two classes, i.e.  
those containing microcalcification clusters and the normal 
tissue\footnote{In 
this paper the tissue not containing 
microcalcification clusters is  referred 
 as normal breast tissue, i.e. in our notation  
this class of tissue can even accommodate 
regions of mammograms
affected by the presence of different pathologies, such as opacities, 
massive lesions, etc.};
\item OUTPUT: merging of contiguous or partially overlapping sub-images and
visualization of the final output
by superimposing rectangles indicating suspicious areas 
to the original image.
\end{itemize}
\section{Tests and results}

\begin{figure}[t]
 \vspace{6.0cm}
\includegraphics{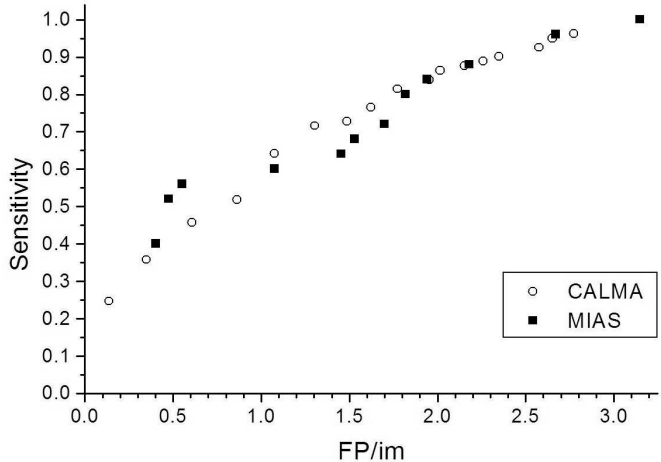}
 \caption{\it FROC curve obtained on the CALMA
dataset (140 mammograms) and on the MIAS
dataset (42 mammograms).
    \label{fig:FROC} }
\end{figure}

The CADe system was set up and tested on a private database of mammograms collected in the framework of  the INFN (Istituto Nazionale di Fisica Nucleare)-founded CALMA (Computer-Assisted Library for MAmmography) project~\cite{database}.
The digitized images are characterized
by a 85$\mu$m pixel pitch and a 12-bit resolution, thus allowing up to 4096 gray levels.
The dataset used for training the CADe
consists of 305 mammograms containing
microcalcification clusters and 540 normal mammograms. The system performances  on a test set of 140 CALMA images (70 with
microcalcification clusters and 70 normal images) have been evaluated in terms of the FROC analysis~\cite{Chakraborty} as shown in fig.~\ref{fig:FROC}.
In particular, as shown in the figure, a sensitivity value of 88\% is obtained at a rate of 2.15 FP/im.

In order to test the generalization capability of the system, we evaluated the CADe performances on the
publicly available MIAS database~\cite{Suckling}. Being the MIAS mammograms characterized by a different pixel
pitch (50$\mu$m instead of 85$\mu$m) and a less deep dynamical range (8 bit per pixel instead of 12) with respect
to the CALMA mammograms, we had to define a tuning procedure for adapting the CADe system to the
MIAS database characteristics. 
A scaling of the wavelet analysis
parameters  allows the
CADe filter to generate very similar pre-processed
images on both datasets. The remaining steps of the analysis,
i.e. the characterization and the classification of
the sub-images, have been directly imported from
the CALMA CADe neural software. The performances
the rescaled CADe achieves on the images
of the MIAS database have been evaluated
on a set of 42 mammograms (20 with microcalcification clusters and 22 normal) and are shown in fig.~\ref{fig:FROC}.
As can be noticed, a sensitivity value of 88\% is obtained at a rate of 2.18 FP/im.

\section{Conclusions}

The implementation of the wavelet transform in the preprocessing step of the analysis and the use of an auto-associative neural network for the automatic feature extraction make our CADe system tunable to different databases.
The main advantage of this procedure is that this scalable CADe system can be tested even on very small databases,  
  i.e. databases not allowing for the learning procedure of the neural networks to be properly carried out.
The strong similarity in the trends of the
FROC curves obtained on the CALMA and on
the MIAS databases provides a clear evidence that
the CADe system we developed can be applied to
different databases with no sensible decrease in the detection performance.


\begin{thebibliography}{99}


\bibitem{database} 
U.~Bottigli  {\it et al}, Search of microcalcification clusters with
the CALMA CAD station, The International Society for Optical Engineering
(SPIE) {\bf 4684} 1301 (2002).
\bibitem{Chakraborty} 
D.~Chakraborty,  Free-response methodology: Alternative analysis and a new observer-performance experiment, Radiology  {\bf 174}(3) 873 (1990).
\bibitem{Suckling} 
J.~Suckling {\it et al},  The mammographic images analysis society digital mammogram database, Excerpta Medica, International Congress Series {\bf 1069} 375 (1994).


%
%
\end{thebibliography}
\end{document}